# Comment on "Atomic Scale Structure and Chemical Composition across Order-Disorder Interfaces"

B. H. Ge and J. Zhu*
Beijing National Center for Electron Microscopy
Department of Materials Science and Engineering
Tsinghua University
Beijing 100084, China

Interfaces have long been known to be the key to many mechanical and electric properties[1]. To nickel base superalloys which have perfect creep and fatigue properties and have been widely used as materials of turbine blades [2], interfaces determine the strengthening capacities in high temperature. By means of high resolution scanning transmission electron microscopy (HRSTEM) and 3D atom probe (3DAP) tomography, Srinivasan *et al.* [3] proposed a new point that in nickel base superalloys there exist two different interfacial widths across the γ/γ´ interface, one corresponding to an order-disorder transition, and the other to the composition transition. We argue about this conclusion in this comment.

Srinivasan *et al.* showed an averaged intensity profile and intensity ratio of a filtered high angle annular dark field (HAADF) image of HRSTEM. However, their original image (the non-filtered image) shown as Fig. 1(b) in reference [3] does not show high quality so that there is too much difference between the non-filtered image and the filtered image as shown in Fig. 1(c) in [3], then their result in [3] is not reliable.

We have done some work in the same way. Figure 1(a) in this paper is a filtered HAADF image of a nickel base superalloy made of Al, Cr, Co, W, Mo, Ta, Re, etc, which was acquired by using a FEI Titan 80-300 microscope. Same as in Ref. [3], the averaged intensity profile across interfaces (corresponding to the area denoted by a rectangle shown in Fig. 1(a)) has been plotted as shown in Fig. 1(b). The higher background intensity in the right side corresponds to higher concentration of heavy alloying elements, vice versa. Then the intensity ratio of each atomic column to its adjacent column in the same area was made displaying in Fig. 1(c), where the ratio in the left side alternates between about 1.1 and 0.9, corresponding to ordered γ´ phases, while in the right side the ratio remains almost constant, i.e. 1, corresponding to the disordered γ phases. Comparing Fig. 1(b) with Fig. 1(c), the transition area of chemical composition is the same as the transition area from ordered phases to disordered phases with the width about 2.2 nm, which denoted by two dotted lines. Note that the same results can be obtained from our non-filtered image.

In conclusion, our result is different from Srinivasan's in [3], which will not be convincing until the same conclusion can be obtained from their non-filtered image (Fig. 1(b) in [3]). However, it is certain in this comment that two different interfacial widths in nickel base superalloys may be not universal.

This work is financially supported by National 973 Project of China and Chinese National Nature Science Foundation. This work made use of the resources of the Beijing National Center for Electron Microscopy.

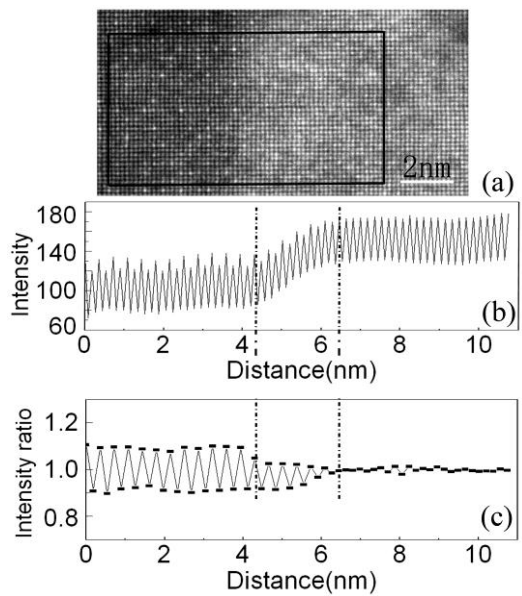

FIG. 1 (a) Filtered HAADF image, (b) averaged intensity profile across interfaces corresponding to the area denoted by a rectangle as shown in Fig. (a), (c) intensity ratios between adjacent columns.